\documentclass[reprint,amsmath,amssymb,aps,prl,superscriptaddress,nofootinbib]{revtex4-2}

\usepackage{graphicx}
\graphicspath{{qnm/}{./}}
\usepackage{bm}
\usepackage{xcolor}
\usepackage{placeins}
\usepackage{chngcntr}
\usepackage[colorlinks=true,citecolor=blue!55!black,
linkcolor=blue!55!black,urlcolor=blue!55!black]{hyperref}

\newcommand{\Om}{\Omega}
\newcommand{\Res}{\operatorname*{Res}}

\newcommand{\ii}{\mathrm{i}}
\newcommand{\dd}{\mathrm{d}}

\begin{document}

\title{Quasinormal-Mode Collider Spectroscopy}

\author{Mian Zhu}
\affiliation{College of Physics, Sichuan University, Chengdu 610065, China}

\author{Jun Tao}
\email{taojun@scu.edu.cn}
\affiliation{College of Physics, Sichuan University, Chengdu 610065, China}

\author{Peng Wang}
\email{pengw@scu.edu.cn}
\affiliation{College of Physics, Sichuan University, Chengdu 610065, China}

\date{\today}

\begin{abstract}
Black-hole (BH) quasinormal modes encode the properties of merger remnants and provide powerful tests of strong-field gravity. The BH spectroscopy has predominantly treated them as independent linear oscillations. Nonlinear ringdown can additionally reveal how these modes interact and which intermediate states mediate their interactions. Inspired by cosmological collider physics, we introduce quasinormal-mode collider spectroscopy (QCS), a framework for organizing and predicting nonlinear BH responses. We establish two structural results. First, at an isolated, nondegenerate intermediate pole, a cubic strain residue factorizes into an excitation amplitude and a quadratic conversion coefficient. This factorization imposes rank constraints across output channels and choices of second-order initial data, providing consistency tests of the intermediate-mode description. Second, the intermediate pole residue can be reconstructed from its horizon and infinity radiation. With the first-order field and third-order data held fixed, this reconstruction extends to the full cubic waveform difference generated by changing the second-order initial data. QCS thus connects lower-order radiation to higher-order observables and provides a framework for probing the interaction structure of nonlinear ringdown.
\end{abstract}

\maketitle

\paragraph{Introduction.---}

The quasinormal mode (QNM) spectrum of a binary-black-hole merger remnant encodes its mass and spin and provides a test of its Kerr nature  \cite{Berti:2009kk,London:2014cma, Cardoso:2016rao,Bhagwat:2019bwv,LIGOScientific:2016aoc,LIGOScientific:2020ibl,KAGRA:2021vkt,LIGOScientific:2025rid,Punturo:2010zz,TianQin:2015yph,Hu:2017mde,LISA:2017pwj,Reitze:2019iox,Yang:2023gas,Cai:2025irl,Chen:2025sbz}. Standard black hole (BH) spectroscopy predominantly treats these modes as independent linear oscillations and therefore does not directly expose the nonlinear interaction vertices of the gravitational field equations. Recent studies of GW250114 have reported evidence for nonlinear spectral components generated by the dominant quadrupolar mode and its overtones \cite{LIGOScientific:2025wao,Yang:2025ror,Wang:2026rev}, adding to growing theoretical and numerical interest in nonlinear ringdown \cite{Mitman:2022qdl,Cheung:2023vki,Ma:2024qcv,Khera:2024bjs,Bourg:2025lpd,Sberna:2021eui,May:2024rrg, Dyer:2025hdt, Emparan:2026lss,Bourg:2026clj,Bucciotti:2026jhr,Keeble:2026epf,Yang:2026eod,Guo:2026dhs}. These developments motivate a systematic understanding of how gravitational interactions shape the nonlinear ringdown spectrum.

Our starting point is the analogy between BH perturbations and cosmological fluctuations. In both cases, the linear mode equations share a one-dimensional wave-equation structure and describe the “free” propagation of perturbations on a fixed background. The interaction terms then couple these free modes, producing nonlinear contributions to BH ringdown and cosmological fluctuations. This viewpoint has already proved fruitful in cosmology. Cosmological collider physics (CCP) has established how primordial correlations encode the masses, spins, and interactions of intermediate particles \cite{Chen:2009we,Chen:2009zp,Arkani-Hamed:2015bza,Xianyu:2022jwk}. Organizing correlations in terms of these particles and their interactions leads to powerful reconstruction principles. An exchanged particle, for example, connects two simpler interactions, and the associated exchange singularity factorizes into lower-order quantities. This permits parts of higher-point correlations to be predicted from simpler interactions \cite{Baumann:2020dch,Baumann:2021fxj,Werth:2024mjg}. Can the same organizing principles reveal the interaction structure of nonlinear BH ringdown?

In this Letter, we introduce quasinormal-mode collider spectroscopy (QCS), a framework that brings the cosmological treatment to nonlinear BH perturbation theory. We demonstrate two structural results through its application to cubic strain in Schwarzschild spacetime. First, at an isolated, nondegenerate intermediate QNM pole, the corresponding cubic strain residue factorizes into an excitation amplitude and a quadratic conversion coefficient, providing a BH counterpart of cosmological exchange factorization \cite{Baumann:2020dch,Baumann:2021fxj}. This factorization imposes rank constraints on the residue matrix across output channels and choices of second-order initial data, providing a consistency test of a single-intermediate-mode description. Second, the intermediate pole residue can be reconstructed from its horizon and infinity radiation. When the first-order field and third-order free data prescription are held fixed, this reconstruction extends to the full cubic waveform difference generated by changing the second-order initial data. Lower-order radiation and quadratic dynamics therefore predict how the cubic response changes with the intermediate excitation. We verify both results in a Schwarzschild example. Together, they establish QCS as a framework for predicting higher-order responses and probing the intermediate modes that mediate nonlinear ringdown.

\begin{figure*}[t]
	\centering
	\includegraphics[width=\textwidth]{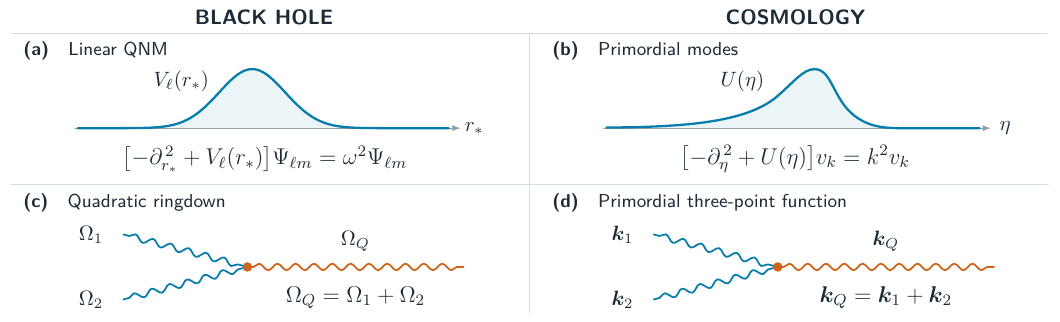}
	\caption{QCS principle and its cosmological analogue. The left column shows black-hole perturbations and the right column cosmological fluctuations. At linear order, (a) the RWZ equation and (b) the Mukhanov--Sasaki equation share a one-dimensional wave-equation structure; (c) two parent modes generate a quadratic ringdown response through a single interaction vertex, while (d) a cubic vertex couples three primordial modes to produce a contact contribution to the three-point function.}
	\label{fig:qcscc}
\end{figure*}

\paragraph{QCS principle.---}

At linear level, the master equations for QNMs and cosmological fluctuations are the Regge--Wheeler--Zerilli (RWZ) equation $[\partial_{r_{\ast}}^2 + \omega^2 - V_l(r)] \Psi_l(\omega,r) = 0$ \cite{Berti:2009kk}, and the Mukhanov--Sasaki equation $\left[\partial_{\tau}^2 +k^2-U(\tau)\right]u_{\bm k}(\tau)=0$ \cite{Mukhanov:1990me}, respectively. Both are one-dimensional wave equations with an effective potential, that is, one-dimensional scattering problems (upper panels of Fig. \ref{fig:qcscc}).

Nonlinear responses arise from interactions among linear excitations, whose propagation is governed by the linearized equations. This structure makes the particle-physics language useful. For example, a contact contribution to a primordial three-point function is represented by a cubic vertex coupling three freely evolving modes (panel (d) of Fig. \ref{fig:qcscc}). A similar organization applies to BH perturbations already at linear \cite{Leaver:1986gd} and nonlinear order \cite{Yang:2015jja, Lagos:2022otp,Iuliano:2024ogr}. A source $J_a(\omega)$ excites a linear strain response $h_a^{(1)}(\omega)$, related schematically by the retarded Green function as $h_a^{(1)}(\omega)=G_a^R(\omega)J_a(\omega)$. When associated with the RWZ operator, $G_a^R$ has poles at the QNM frequencies. Near an isolated simple pole,
\begin{equation}
    \label{eq:Green1}
	G_a^R(\omega) = \frac{Z_A}{\omega-\Om_A} + G_{a,\mathrm{reg}}^R ~,~ Z_A \equiv \Res_{\omega = \Om_A} G_a^R(\omega) ~,
\end{equation}
where $Z_A$ is the pole residue and $G_{a,\mathrm{reg}}^R$ is analytic near $\Omega_A$. 

This viewpoint becomes particularly useful at nonlinear order, where interactions between linear modes generate additional components of the ringdown signal. For instance, the quadratic response in panel (c) of Fig. \ref{fig:qcscc} arises from two parent modes coupled through a single interaction vertex. Its amplitude can be organized as
\begin{equation}
\label{eq:AQfactor}
    \mathcal A_Q^{(2)}(t_*) = \mathcal C_{Q|BC}(t_*) \mathcal A_B^{(1)}(t_*)\mathcal A_C^{(1)}(t_*) ~,
\end{equation}
with the interaction vertex, output propagation, angular factors, and nonlinear strain extraction absorbed into $\mathcal C_{Q|BC}$. For the familiar Schwarzschild self-coupling $220\times220$, this reproduces the known relation $\mathcal A_Q^{(2)}\propto[\mathcal A_{220}^{(1)}]^2$, where $Q$ denotes the quadratic daughter in the $(4,4)$ angular channel \cite{Bucciotti:2024zyp, Bucciotti:2024jrv}. On a stationary background, the quadratic source oscillates at the sum of the parent frequencies, giving the driven daughter frequency $\Omega_Q=2\Omega_{220}$ \cite{Ioka:2007ak,Nakano:2007cj}. This frequency addition reflects time-translation symmetry, analogous to momentum conservation $\bm k_Q=\bm k_1+\bm k_2$ arising from spatial-translation symmetry in cosmology.

Now, organizing the amplitudes as in Eq. \eqref{eq:AQfactor} gives a simple consequence. Since each mode amplitude transforms as $\mathcal A_X(t + \delta t)=\mathcal A_X(t)e^{-i\Omega_X\delta t}$, the ratio $\mathcal C_X(t)\equiv\mathcal A_X(t)/[\mathcal A_{220}^{(1)}(t)]^2$ is invariant under reference-epoch shifts for the quadratic daughter. An independently excited linear $440$ mode instead produces drift in both magnitude and phase at a rate fixed by the known frequency mismatch $\Omega_{440} \neq 2\Omega_{220}$ (Appendix~\ref{sec:quadratic-response}). The organization principle thus turns the familiar Schwarzschild result into a reference-epoch invariance, supplying a consistency condition for the sum-frequency assignment.

\begin{figure*}[t]
    \centering
 \includegraphics[width=0.9\linewidth]{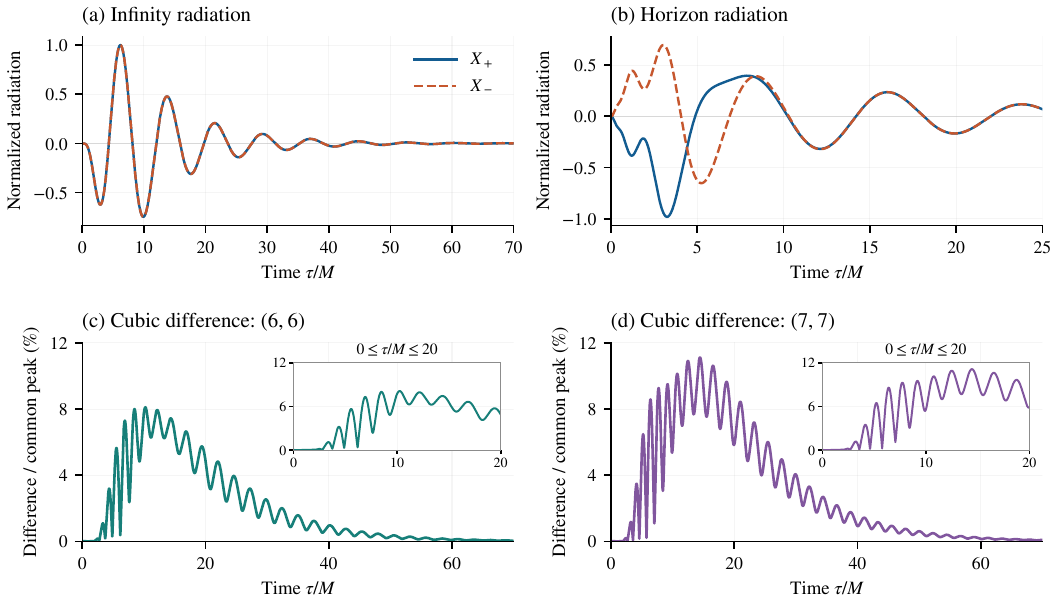}
 \caption{Preparation information revealed by cubic radiation. Two Schwarzschild preparations $X_\pm$, with the same first-order field and cubic-data prescription, produce nearly identical second-order radiation at infinity (a) but distinct horizon radiation (b). Their normalized cubic waveform difference $100|\delta h_{\ell m}^{(3)}|/H_{\ell m}$, however, can differ approximately 10\% for the $(6,6)$ (c) and $(7,7)$ (d) channels. Here $\delta h_{\ell m}^{(3)} =h_{\ell m,X_+}^{(3)}-h_{\ell m,X_-}^{(3)}$ and $H_{\ell m}=\max_{\tau,s=\pm}|h_{\ell m,X_s}^{(3)}|$. Relevant details are organized in Appendix~\ref{secS:numerics}.}
 \label{fig:prep}
\end{figure*}

\begin{figure}[t]
    \centering
    \includegraphics[width=\columnwidth]{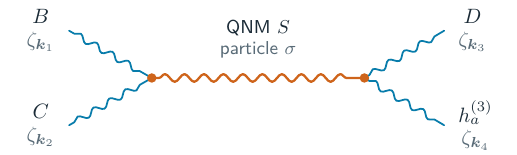}
    \caption{The exchange contribution in QCS (upper labels) and CCP (lower labels). }
    \label{fig:exchange}
\end{figure}

\begin{figure*}[t]
    \centering
 \includegraphics[width=\linewidth]{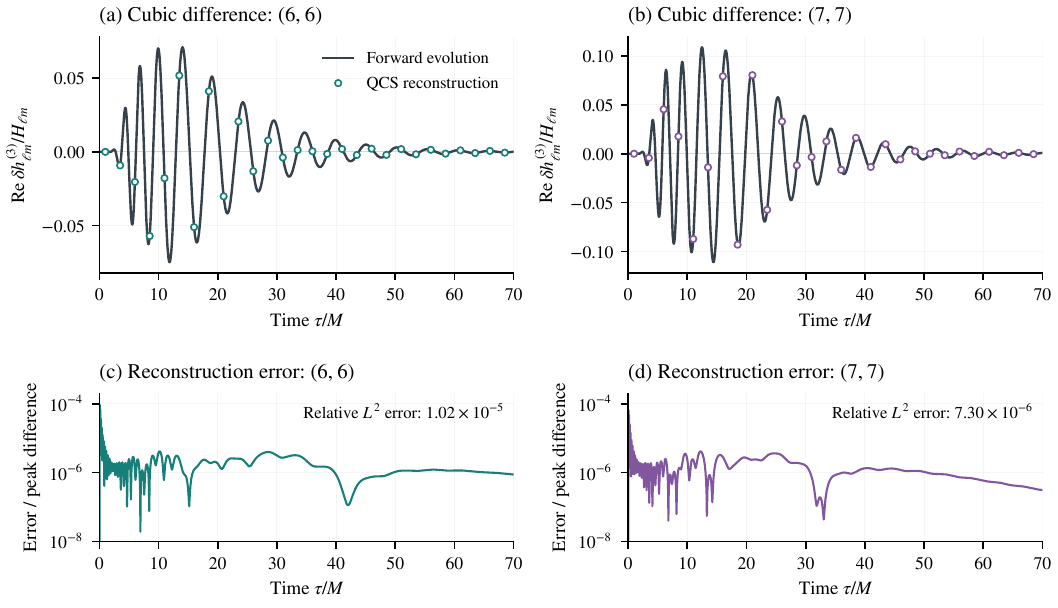}
 \caption{Reconstruction of cubic waveform differences from lower-order radiation. Panels (a,b) compare the forward-evolved difference $\delta h_{\ell m}^{(3)}$ (lines) with its QCS reconstruction (circles) in the $(6,6)$ and $(7,7)$ channels for the preparations in Fig. \ref{fig:prep}. Panels (c,d) show the complex reconstruction error divided by $\max_\tau|\delta h_{\ell m}^{(3)}|$ in the corresponding channel. The relative complex $L^2$ errors over $0\leq\tau/M\leq70$ are $1.02\times10^{-5}$ and $7.30\times10^{-6}$, respectively. These noiseless comparisons verify the prediction of the preparation-dependent cubic response from lower-order radiation (Appendix~\ref{secS:numerics}).}
 \label{fig:reconstruction}
\end{figure*}

There remains, however, a potential challenge: the quadratic record at infinity does not fully determine the the prepared second-order field. The parents can determine only the sourced part of the second-order response, thus two preparations (i.e., distinct second-order initial data for the same parents) can differ by a freely propagating perturbation. Now, complete future scattering data contain both horizon and infinity channels \cite{Xuan:2020wif,Masaood:2022bvi}, but a distant observer records only the latter, so two preparations differed at horizon direction can leave nearly identical quadratic records at infinity. We exhibit this degeneracy quantitatively in Appendix~\ref{secS:numerics} and demostrate a Schwartzchild case in Fig. \ref{fig:prep}. Therefore, distinguishing them requires probing the daughter beyond its own quadratic radiation at infinity.

\paragraph{Exchange spectrum and its structure.---}

The aforementioned ambiguity motivates going beyond quadratic order. Specifically, the second-order daughter excited by parents $B$, $C$ can interact further with a first-order spectator $D$, generating the cubic strain $h_a^{(3)}$ (Fig. \ref{fig:exchange}). This interaction transfers the hidden preparation difference into cubic radiation, so cubic responses can distinguish preparations that the quadratic infinity record cannot. The analogous process in CCP is the exchange diagram, where an intermediate particle $\sigma$ connects two interactions contributing to a primordial four-point function; we accordingly term this nonlinear ringdown process ``exchange.''

We then illustrate how the exchange contribution can reproduce the information of the intermediate QNMs. In the one-sided Laplace variable $s=-\ii\omega$, the second-order master response to parents $B,C$ reads
\begin{equation}
    \label{eq:prepared-response}
	p_\beta(s) = \mathcal G^R(s) f_\beta(s)~,~ \mathcal G^R(s) = \mathcal L(s)^{-1} ~,
\end{equation}
where parent amplitudes are removed. Here $f_\beta$ contains the quadratic source and initial-data terms for causal preparation $\beta$ (Appendix~\ref{secS:prepared-exchange}). We see that for a fixed spectator QNM (D), the preparation-dependent cubic response depends linearly on the intermediate-field difference. This interaction can therefore convert preparation differences that are weakly visible in quadratic radiation at infinity into distinguishable cubic signals.

We now derive a structural constraint on cubic strain across output channels and second-order preparations. At a simple, nondegenerate intermediate pole isolated from other response singularities, the normalized intermediate profile is fixed, while its excitation depends on the preparation \cite{Green:2022htq}. Applying the quadratic conversion to this profile yields the factorization (Appendix~\ref{secS:prepared-exchange})
\begin{equation}
    \label{eq:factorization}
	K_{\alpha\beta}^{(S)} = \mathcal V_{\alpha S}\,\mathcal E_{S\beta}~.
\end{equation}
Here $K_{\alpha\beta}^{(S)}$ is the residue in the output Laplace variable at $s_S+s_D$, equivalently the coefficient of $e^{-i(\Omega_S+\Omega_D)u}$, with parent and spectator amplitudes removed. The index $\alpha=(a,D)$ specifies the output and spectator channel. The excitation $\mathcal E_{S\beta}$ carries the preparation dependence, while $\mathcal V_{\alpha S}$ includes quadratic conversion, output propagation, and nonlinear strain extraction (Appendix~\ref{secS:prepared-exchange}).

The factorization \eqref{eq:factorization} has an immediate consequence. The pole fixes the intermediate profile, so changing the preparation rescales every output by the same factor. The coefficients $K_{\alpha\beta}^{(S)}$ in Eq. \eqref{eq:factorization} form a matrix whose rows label output channels and columns label preparations. Its columns are therefore proportional, giving rank at most one. A sum of $N$ such intermediate contributions has rank at most $N$. Physically, the rank counts the independent output patterns resolved across preparations and thus gives a lower bound on the number of intermediate contributions. This upper bound  is essentially a consistency relation that can be used to test the single-pole description.

To make the test quantitative, define the cross ratio $z\equiv (K_{\alpha\beta}K_{\alpha'\beta'})/ (K_{\alpha\beta'} K_{\alpha'\beta}) $ of any two outputs $\alpha,\alpha'$ and two preparations $\beta,\beta'$. For a single intermediate mode, $K$ is an outer product, so its factors cancel pairwise and $z=1$ We thus introduce the bounded statistic $\epsilon_{\rm rk}\equiv2|z-1|/(|z|+1)$, which vanishes exactly at rank one. For a cubic strain containing two exchange contributions associated with distinct intermediate poles, the corresponding exchange-coefficient matrix is a sum of outer products and generically has rank two. Consequently, $\epsilon_{\rm rk}$ is generically nonzero. This is illustrated with a two-pole model in Appendix~\ref{secS:two-pole} (with parent pairs playing the role of preparations). A nonzero value therefore falsifies the single-intermediate-mode description of the extracted exchange matrix. We verify the single-pole factorization directly in Schwarzschild: for the $440$-mediated exchange in the $(6,6)$ and $(7,7)$ channels across two preparations, forward evolution reproduces the factorized residues within $2.6\times10^{-10}$, and the coefficient matrix has singular-value ratio $9\times10^{-15}$ (Appendix~\ref{secS:numerics}).

\paragraph{Reconstruction of cubic strain.---}

The factorization \eqref{eq:factorization} reduces the prediction of cubic coefficients to the determination of conversion factor $\mathcal V_{\alpha S}$ and excitation $\mathcal E_{S\beta}$. Now,  the factor $\mathcal V_{\alpha S}$ follows from the quadratic interaction of the second-order intermediate perturbation with the first-order spectator, which contributes at cubic order. With their normalized profiles fixed by linear theory, this conversion is determined by quadratic gravitational dynamics. The remaining question now is to determine the preparation-dependent excitation $\mathcal E_{S\beta}$. The recovery of this excitation from the intermediate mode's second-order radiation provides a classical counterpart of cosmological exchange reconstruction from lower-point ingredients \cite{Werth:2024mjg}. In particular, at an isolated pole, a linear scattering identity reconstructs the intermediate residue from the horizon and infinity radiation residues of the prepared second-order response (Appendix~\ref{secS:reconstruction}). Applying the quadratic conversion to this reconstructed residue then determines $K_{\alpha\beta}^{(S)}$ from lower-order radiation and dynamics.

Interestingly, a comparison of different preparations can extend this reconstruction from isolated pole coefficients to full time-dependent cubic waveform differences. We compare two preparations that differ only in their second-order initial data. Let $\phi^{(n)}$ denote the order-$n$ metric perturbation with tensor indices suppressed. The second-order master-field difference $q$ propagates freely. For each preparation $X$, let $\Psi_{b,X}^{(2)}$ denote its second-order master-field radiation history at the horizon ($b=H$) or infinity ($b=\infty$). Their differences reconstruct the intermediate field through \cite{Masaood:2022bvi,Xuan:2020wif}
\begin{equation}
\Delta\Psi_b^{(2)} =\Psi_{b,X_+}^{(2)}-\Psi_{b,X_-}^{(2)} ~,~ q =\mathcal I_+ [\Delta\Psi_H^{(2)},\Delta\Psi_\infty^{(2)}] ~.
\label{eq:field-reconstruction}
\end{equation}
Here $\mathcal I_+$ is the linear scattering reconstruction map.  Subtracting the third-order equations cancels the common genuine cubic source. The purely first-order cubic contribution to strain also cancels, leaving
\begin{equation}
\Delta h_\alpha^{(3)} = P_\alpha G_\alpha^R B_\alpha(\phi^{(1)},\Delta\phi^{(2)}) + Q_\alpha(\phi^{(1)},\Delta\phi^{(2)}) ~.
\label{eq:rec}
\end{equation}
In the reconstruction process, the bilinear map $B_\alpha$ couples the reconstructed difference to the common first-order metric perturbation $\phi^{(1)}$; $G_\alpha^R$ propagates the resulting source, and $P_\alpha$ extracts its linear strain. The bilinear map $Q_\alpha$ supplies the nonlinear strain contribution. Equations \eqref{eq:field-reconstruction} and \eqref{eq:rec} therefore determine the complete cubic strain difference from lower-order radiation and quadratic gravitational dynamics. This reconstruction includes the transient response and requires no isolated intermediate pole (Appendix~\ref{secS:reconstruction}).

We verify the reconstruction Eq. \eqref{eq:rec} by evaluating the relative complex $L^2$ errors in cubic channels considered in Fig. \ref{fig:prep}, and presenting the result in Fig. \ref{fig:reconstruction}. The relative errors are of order $10^{-5}$ in both cubic channels, and details can be found in Appendix~\ref{secS:numerics}. Therefore, our reconstruction process turns the preparation dependence exhibited in Fig. \ref{fig:prep} into a quantitative prediction for the amplitude and time evolution of the cubic waveform difference, which is composed by lower-order boundary radiation and quadratic dynamics. 

\paragraph{Conclusions.---}
In this Letter, we proposed quasinormal-mode collider spectroscopy (QCS), a systematic framework that organizes nonlinear BH perturbations through propagating modes and their interactions. We apply this methodology in an exchange contribution, by separating local contact responses from pole-mediated exchange, thereby isolating the physics of intermediate QNMs in nonlinear ringdown. Analogue to the CCP, we find a factorization of cubic responses in exchange contributions, and an associated rank condition as consistency relations. We also reconstruct complete cubic waveform differences from lower-order horizon and infinity radiation, with the first-order field and third-order free data held fixed.

The natural next step is to extend our framework to Kerr cases. A preliminary analysis in Appendix~\ref{secS:Kerr} indicates that the combined linear pole factor remains finite in near-extreme Kerr cases, while the known zero-damping overtone tower \cite{Yang:2012pj,Berti:2006wq} becomes closely spaced enough that a single intermediate mode is not parametrically isolated. These findings motivate a collective treatment of intermediate poles in the QCS framework using existing projection and nonlinear perturbation methods \cite{Green:2019nam,Spiers:2023cip}. Computing the corresponding nonlinear Kerr couplings would turn the structural relations developed here into quantitative, falsifiable predictions for astrophysical BHs.

Observationally, QCS provides targets for future gravitational-wave measurements. The reference-epoch invariance of the ratio $\mathcal{C}_X$ for a quadratically driven daughter is the most immediate consistency check, whereas exchange-rank relations require several nonlinear channels and greater sensitivity. These developments  would extend BH spectroscopy from identifying resonances to probing their interactions and the intermediate modes that mediate them.

\paragraph{Acknowledgements.---}
We thank Yi-Fu Cai, Wei Hong, Hyat Huang, Yuxiao Liu, Xin Ren, Yi Wang, and Yuhang Zhu for helpful discussions. This work is supported by the National Natural Science Foundation of China
(NSFC) with Grants Nos.12175212, 12275183, 12503005, Sichuan Science and Technology Program Grant No. 2026NSFSC0804, and the Fundamental Research Funds for the Central Universities Grant No. YJ202551. We acknowledge the usage of the following datasets: the Kerr quasinormal-mode frequencies and excitation factors used in this Letter are publicly available in the dataset of Motohashi \cite{MotohashiQNEFData}. The quadratic-response code and data are available in the QuadraticQNM repository \cite{BucciottiQuadraticQNMCode}. 

\appendix
\counterwithout{equation}{section}
\setcounter{secnumdepth}{2}
\setcounter{equation}{0}
\setcounter{figure}{0}
\setcounter{table}{0}
\renewcommand{\theequation}{S\arabic{equation}}
\renewcommand{\thefigure}{S\arabic{figure}}
\renewcommand{\thetable}{S\arabic{table}}

\section{Quadratic response and amplitude conventions}
\label{sec:quadratic-response}

\subsection{Perturbative equations and physical strain}

Consider a metric expanded about a stationary black-hole background,
\begin{equation}
 g_{\mu\nu}=\bar g_{\mu\nu}+\epsilon\phi^{(1)}_{\mu\nu}
 +\epsilon^2\phi^{(2)}_{\mu\nu}+\epsilon^3\phi^{(3)}_{\mu\nu}+\cdots~.
 \label{eqS:metric-expansion}
\end{equation}
We use $\phi^{(n)}$ for the metric perturbation, with tensor indices suppressed below, and $h^{(n)}_a$ for its physical strain in angular channel $a$. The expansion contains no factorials. Let $\mathcal E[g]=0$ denote the gravitational field equations and define their Taylor maps by
\begin{equation}
 \mathcal L=\mathcal E^{(1)}~,~
 B=-\mathcal E^{(2)}~,~ C=-\mathcal E^{(3)}~.
\end{equation}
In a fixed gauge, the perturbative equations in the vacuum evolution region are
\begin{align}
 \mathcal L\phi^{(1)}&=0~,\nonumber\\
 \mathcal L\phi^{(2)}&=\tfrac12 B(\phi^{(1)},\phi^{(1)})~,\nonumber\\
 \mathcal L\phi^{(3)}&=B(\phi^{(1)},\phi^{(2)})
 +\tfrac16 C(\phi^{(1)},\phi^{(1)},\phi^{(1)})~.
 \label{eqS:perturbative-equations}
\end{align}
Thus the same linearized operator propagates every perturbative order. The lower-order fields enter through the sources and the prescribed initial data.

The observable has a corresponding expansion. Writing its linear, quadratic, and cubic Taylor maps as $P_a,Q_a,T_a$, respectively, gives
\begin{align}
 h_a^{(2)}&=P_a\phi^{(2)}+\tfrac12 Q_a(\phi^{(1)},\phi^{(1)})~,\nonumber\\
 h_a^{(3)}&=P_a\phi^{(3)}+Q_a(\phi^{(1)},\phi^{(2)})
 +\tfrac16 T_a(\phi^{(1)},\phi^{(1)},\phi^{(1)})~.
 \label{eqS:observable-expansion}
\end{align}
These expressions distinguish a master-field response from the measured strain. Both propagation and nonlinear strain extraction enter the conversion coefficients used in the main text.

To see how frequency addition arises, write the quadratic source schematically as
\begin{align}
 S_a^{(2)}(\omega)=\sum_{bc}\int\frac{\dd\omega_1}{2\pi}
 \frac{\dd\omega_2}{2\pi}\,&2\pi\delta(\omega-\omega_1-\omega_2)
 \nonumber\\[-2pt]
 &\times\Gamma^{(2)}_{a|bc}\phi_b^{(1)}(\omega_1)
 \phi_c^{(1)}(\omega_2)~.
 \label{eqS:quadratic-source}
\end{align}
Spatial contractions and derivatives are included in $\Gamma^{(2)}$. Stationarity fixes the frequency delta function. The sourced master response is $G_a^RS_a^{(2)}$, supplemented by its initial-data solution; Eq.~\eqref{eqS:observable-expansion} then gives the strain.

We use $\widetilde f(\omega)=\int\dd t\,e^{\ii\omega t}f(t)$. A mode switched on at $t_*$ has transform
\begin{equation}
 \widetilde h_X(\omega)=
 \frac{\ii\mathcal A_X(t_*)e^{\ii\omega t_*}}{\omega-\Om_X}~.
 \label{eqS:fourier-mode}
\end{equation}
The retarded convolution of two parent poles satisfies
\begin{equation}
 \int\frac{\dd\omega_1}{2\pi}
 \frac{1}{(\omega_1-\Om_B)(\omega-\omega_1-\Om_C)}
 =\frac{-\ii}{\omega-\Om_B-\Om_C}~.
\end{equation}
Closing the retarded contour below the real axis gives a daughter at $\Om_Q=\Om_B+\Om_C$, with time-domain amplitude
\begin{equation}
 \mathcal A_Q^{(2)}(t_*)=
 \mathcal C_{Q|BC}\mathcal A_B^{(1)}(t_*)\mathcal A_C^{(1)}(t_*)~,
 \label{eqS:quadratic-amplitude}
\end{equation}
where $\mathcal C_{Q|BC}$ contains the source, output propagation, angular projection, and strain extraction. The Fourier residue is $\ii e^{\ii\Om_Qt_*}\mathcal A_Q^{(2)}$. In the one-sided Laplace variable $s=-\ii\omega$, the residue equals the coefficient of $e^{s\tau}$.

\subsection{Schwarzschild self-coupling}
\label{secS:RWZ-benchmark}

The $220\times220$ coupling provides a normalization check for Eq.~\eqref{eqS:quadratic-amplitude}. Here $\ell mn$ labels a linear QNM. The quadratic daughter $Q$ has angular indices $(4,4)$ and frequency $2\Om_{220}$; $440$ denotes the independent linear fundamental.

At infinity, we define the strain by
\begin{equation}
 h_+-\ii h_\times=\frac{M}{r}\sum_{\ell m}
 h_{\ell m}(u)\,{}_{-2}Y_{\ell m}~,~ u=t-r_*~.
 \label{eqS:strain-convention}
\end{equation}
The regularized even-parity quadratic field obeys
\begin{equation}
 \left[\frac{\dd^2}{\dd r_*^2}+\Om_Q^2-V_4^+\right]\Psi_4^{(2)}
 =\mathfrak S^{EEE}_{4|22,22}~,
 \label{eqS:quadratic-Zerilli}
\end{equation}
where $\dd r_*/\dd r=f^{-1}$, $f=1-2M/r$, and
\begin{align}
 V_\ell^+&=f\frac{2\lambda^2(\lambda+1)r^3+6\lambda^2Mr^2
 +18\lambda M^2r+18M^3}{r^3(\lambda r+3M)^2}~,\nonumber\\
 \lambda&=\frac{(\ell-1)(\ell+2)}2~.
 \label{eqS:Zerilli-potential}
\end{align}
We use the regularized source and asymptotic strain reconstruction of Refs.~\cite{Bucciotti:2024zyp,Bucciotti:2024jrv,BucciottiQuadraticQNMCode}; $EEE$ specifies even parity for both parents and the output. The outgoing master amplitude, divided by the two parent master amplitudes, defines $\mathcal I_{4|22,22}$. It is evaluated by a Leaver-series solution with the QNM boundary factors extracted, implementing the analytic continuation of the retarded response.

Including inverse regularization and nonlinear strain reconstruction gives
\begin{equation}
 \widehat{\mathcal C}_{4|22,22}
 =\frac{\sqrt{10}}2\mathcal I_{4|22,22}+\mathcal C_{\rm rec}~.
 \label{eqS:RWZ-conversion}
\end{equation}
The $N=80$ calculation, with $M\Om_{220}\simeq0.37367-0.088962\ii$, yields $\widehat{\mathcal C}_{4|22,22}\simeq0.9187-0.06266\ii$, consistent with the published reduced coefficient $0.919-0.0627\ii$. Restoring the identical-parent factor and angular projection gives
\begin{align}
 \mathcal C_{Q|220,220}
 &=\frac12\begin{pmatrix}4&2&2\\-4&2&2\end{pmatrix}
 \widehat{\mathcal C}_{4|22,22}\nonumber\\
 &=\frac16\widehat{\mathcal C}_{4|22,22}
 \simeq0.1535e^{-0.06810\ii}~.
 \label{eqS:benchmark}
\end{align}
The factor $1/6$ distinguishes the physical strain ratio from the reduced coefficient in Eq.~\eqref{eqS:RWZ-conversion}.

Now, for $h_X(t)=\mathcal A_X(t_*)e^{-\ii\Om_X(t-t_*)}$, an epoch shift gives $\mathcal A_X(t_*+\delta t)=\mathcal A_X(t_*)e^{-\ii\Om_X\delta t}$. The ratio $\mathcal C_X=\mathcal A_X/[\mathcal A_{220}^{(1)}]^2$ therefore transforms as
\begin{equation}
 \mathcal C_X(t_*+\delta t)=\mathcal C_X(t_*)
 e^{-\ii(\Om_X-2\Om_{220})\delta t}~.
 \label{eqS:epoch-transport}
\end{equation}
The quadratic daughter has constant magnitude and phase. A linear $440$ component instead has $M(\Om_{440}-2\Om_{220})\simeq0.06184+0.08376\ii$. That means, shifting the epoch by $10M$ multiplies the ratio's magnitude by $2.31$ and changes its phase by $-0.618$ rad, therefore distinguishable from the $220 \times 220 \to 440$ quadratic QNMs.

\section{Prepared exchange and pole factorization}
\label{secS:prepared-exchange}

\subsection{The causal intermediate response}

An exchange contribution consists of two quadratic interactions connected by a propagating intermediate field. For a routing $B+C\to S$, followed by $S+D\to a$, its frequency-space kernel contains
\begin{equation}
 G_a^R\Gamma^{(2)}_{a|SD}G_S^R\Gamma^{(2)}_{S|BC}~.
\end{equation}
The intermediate propagator supplies the exchange singularity. To connect this organization to a causal waveform, we also specify how the intermediate field is prepared.

Let $\tau=0$ be the initial slice and remove the parent amplitudes from the second-order master field. For an evolution equation
\begin{equation}
 \ddot p=L_1p+L_2\dot p+b(\tau)~,
\end{equation}
define the one-sided Laplace transform $p_\beta(s)=\int_0^\infty\dd\tau\,e^{-s\tau}p_\beta(\tau)$. A preparation $\beta$ specifies $p_{0\beta}=p_\beta(0)$ and $v_{0\beta}=\dot p_\beta(0)$. Transforming the time derivatives gives
\begin{align}
 \mathcal L(s)&=s^2-sL_2-L_1~,\nonumber\\
 p_\beta(s)&=\mathcal G^R(s)f_\beta(s)~,~
 \mathcal G^R(s)=\mathcal L(s)^{-1}~,\nonumber\\
 f_\beta(s)&=\widetilde b(s)+(s-L_2)p_{0\beta}+v_{0\beta}~.
 \label{eqS:prepared-source}
\end{align}
This is the prepared response of Eq.~\eqref{eq:prepared-response}. For exponential parents, $b(\tau)=b\,e^{s_p\tau}$ with $s_p=s_B+s_C$, so $\widetilde b=b/(s-s_p)$. The source fixes the driven response, while the initial-data terms fix its homogeneous completion.

Two preparations with the same first-order parents have the same quadratic source. Subtracting their equations therefore leaves a freely propagating second-order difference. This freedom is the origin of the preparation dependence studied below.

\subsection{Factorization of the physical strain residue}
\label{secS:factorization}

Consider an isolated simple pole $s_S=-\ii\Om_S$ of the retarded resolvent. Let $r_S$ and $\ell_S$ be its right and left null vectors, respectively, with
\begin{equation}
 \mathcal L(s_S)r_S=0~,~ \ell_S\mathcal L(s_S)=0~.
\end{equation}
The left vector denotes the analytic dual pairing. For a nondegenerate pole with $\ell_S\mathcal L'(s_S)r_S\ne0$, the singular part of the resolvent is
\begin{equation}
 \mathcal G^R(s)=
 \frac{r_S\otimes\ell_S}{(s-s_S)\ell_S\mathcal L'(s_S)r_S}
 +\mathcal G^R_{\rm reg}(s)~.
 \label{eqS:resolvent-dyad}
\end{equation}
When $f_\beta$ is analytic at $s_S$, Eq.~\eqref{eqS:prepared-source} gives
\begin{align}
 \Res_{s=s_S}p_\beta(s)&=r_S\mathcal E_{S\beta}~,\nonumber\\
 \mathcal E_{S\beta}&=
 \frac{\ell_S f_\beta(s_S)}{\ell_S\mathcal L'(s_S)r_S}~.
 \label{eqS:excitation}
\end{align}
The pole therefore fixes the radial profile, and preparation changes its scalar excitation.

Now hold a normalized spectator $D$ fixed. Its factor $e^{s_D\tau}$ shifts the intermediate singularity to the output frequency $z_S=s_S+s_D$. Let $\mathcal D_\alpha(z)$ denote the conversion of an intermediate profile into strain, with $\alpha=(a,D)$. This map includes the quadratic source, the retarded output propagator, and the quadratic strain term in Eq.~\eqref{eqS:observable-expansion}. Evaluating its analytic part at $z_S$ yields
\begin{align}
 K_{\alpha\beta}^{(S)}
 &\equiv\Res_{z=z_S}\widetilde h_{\alpha\beta}^{(3)}(z)
 =\mathcal V_{\alpha S}\mathcal E_{S\beta}~,\nonumber\\
 \mathcal V_{\alpha S}&=\mathcal D_\alpha(z_S)r_S~.
 \label{eqS:factorization}
\end{align}
Parent and spectator amplitudes are removed in this definition, as in the main text. Since $z$ is a Laplace variable, $K_{\alpha\beta}^{(S)}$ is also the coefficient of $e^{(s_S+s_D)\tau}$.

The required isolation is local: the prepared source, output propagation, strain reconstruction, and other interaction routes are analytic at the selected singularity. Under these conditions, analytic contact terms have zero residue there. Coincident poles or a degenerate intermediate eigenspace require retaining the corresponding larger singular subspace.

\subsection{Rank constraints across channels and preparations}
\label{secS:rank}

Equation~\eqref{eqS:factorization} makes the residue matrix an outer product. Every nonzero column has the same output pattern, and hence
\begin{equation}
 \operatorname{rank}K^{(S)}\leq1~,~
 \operatorname{rank}\!\left(\sum_{j=1}^{N}K^{(S_j)}\right)\leq N~.
 \label{eqS:rank-bound}
\end{equation}
The second bound is saturated when both the conversion vectors and the excitation vectors are linearly independent across the retained modes. The measured rank thus gives a lower bound on the number of independent intermediate contributions resolved by the channel set.

For two outputs and two preparations, write $K_{ij}=K_{\alpha_i\beta_j}$. If $K_{12}K_{21}\ne0$, the cross ratio is
\begin{equation}
 z=\frac{K_{11}K_{22}}{K_{12}K_{21}}~,~
 \epsilon_{\rm rk}=\frac{2|z-1|}{|z|+1}~.
 \label{eqS:rank-statistic}
\end{equation}
Equivalently,
\begin{equation}
 \epsilon_{\rm rk}=
 \frac{2|K_{11}K_{22}-K_{12}K_{21}|}
 {|K_{11}K_{22}|+|K_{12}K_{21}|}~.
 \label{eqS:rank-minor}
\end{equation}
The latter expression remains usable when one product vanishes, provided the denominator is nonzero. It lies between zero and two, and vanishes for a single nonzero outer product. For a larger matrix, the rank-one condition requires all $2\times2$ minors to vanish.

This test also removes independent row and column gains. If fitted amplitudes satisfy $A_{\alpha\beta}=g_\alpha h_\beta K_{\alpha\beta}$ with nonzero gains, both products in Eq.~\eqref{eqS:rank-minor} acquire the same factor $g_1g_2h_1h_2$. The factor cancels from $\epsilon_{\rm rk}$. The physical conversion and excitation coefficients separately retain the normalization freedom
$\mathcal V_{\alpha S}\to c_S\mathcal V_{\alpha S}$,
$\mathcal E_{S\beta}\to c_S^{-1}\mathcal E_{S\beta}$.
Their product is fixed by the strain and preparation conventions.

\section{Reconstruction from lower-order radiation}
\label{secS:reconstruction}

\subsection{Cosmological exchange reconstruction}
\label{secS:cosmological-reconstruction}

To connect the pole factorization of Eq.~\eqref{eqS:factorization} with reconstruction from radiation, we first apply the cosmological exchange-reconstruction strategy to the linear black-hole Green kernel. In cosmology, this strategy joins lower-point ingredients through an internal propagator and uses its analytic structure to select the exchanged contribution \cite{Baumann:2020dch,Baumann:2021fxj,Werth:2024mjg}. Here we derive the propagator's decomposition into radiation channels and continue it to a QNM pole. This provides the linear ingredient for the latter reconstruction process.

Consider the canonical master operator $\mathsf L_\omega=-\partial_x^2+V_\ell-\omega^2$, with $x=r_*$. Its source convention is $\mathsf L_\omega\psi=j$, so $j=-S$ relative to the radial RWZ equation in Sec.~\ref{sec:quadratic-response}. Let $G^R$ and $G^A$ denote its retarded and advanced kernels. For real $\omega>0$ and smooth compact sources $j_i$, the retarded fields $\psi_i=G^Rj_i$ have amplitudes $A_{Hi}$ multiplying $e^{-\ii\omega x}$ at the horizon and $A_{\infty i}$ multiplying $e^{\ii\omega x}$ at infinity. Green's identity gives
\begin{align}
 \int\dd x\,(j_i^*\psi_j-\psi_i^*j_j)
 &=[\psi_i^*\psi_j'-(\psi_i^*)'\psi_j]_{-\infty}^{+\infty}
 \nonumber\\
 &=2\ii\omega\sum_{b=H,\infty}A_{bi}^*A_{bj}~.
 \label{eqS:flux-boundary}
\end{align}
Define the boundary kernels by $A_b[j]=\int\dd y\,K_b^R(y,\omega)j(y)$. Since $G^A=(G^R)^\dagger$ on the real axis, arbitrary source pairs yield
\begin{equation}
 G^R-G^A=2\ii\omega\sum_{b=H,\infty}K_b^A\otimes K_b^R~,
 \label{eqS:scattering-identity}
\end{equation}
where $K_b^A=(K_b^R)^*$ at real frequency. The two terms are the horizon and infinity radiation channels. Their normalization follows from the boundary flux, including the radiation conditions of the Green functions.

To exhibit the exchange operation, write a response schematically as $\mathcal H_{\alpha\beta}=D_\alpha G^Rj_\beta+\mathcal A_{\alpha\beta}$, with the spectator frequency shift understood. Here $j_\beta$ describes production, $D_\alpha$ converts the intermediate field into the output, and $\mathcal A_{\alpha\beta}$ is analytic at the selected pole. For compact sources and regular output functionals, replacing the specified internal line gives
\begin{align}
 \Delta_{\rm int}\mathcal H_{\alpha\beta}
 &\equiv D_\alpha(G^R-G^A)j_\beta\nonumber\\
 &=2\ii\omega\sum_{b=H,\infty}
 D_\alpha[K_b^A]\,A_b[j_\beta]~.
 \label{eqS:exchange-channel-operation}
\end{align}
This expresses the selected spectral contribution through production and conversion factors, with the external ingredients held fixed.

Continue the scattering kernels analytically to an isolated, simple, nondegenerate retarded pole $\omega=\Om_S$, where $G^A$ is regular. Taking the residue of Eq.~\eqref{eqS:scattering-identity} gives
\begin{equation}
 \Res_{\omega=\Om_S}G^R
 =2\ii\Om_S\sum_{b=H,\infty}K_b^A(\Om_S)
 \otimes\Res_{\omega=\Om_S}K_b^R~.
 \label{eqS:spectral-residue}
\end{equation}
For production and conversion analytic at this pole, the internal-line replacement preserves the exchange residue. Its one-dimensional intermediate profile gives the excitation--conversion factorization of Eq.~\eqref{eqS:factorization}. This is the classical spectral construction used here; quantum cosmological cutting rules additionally involve state-dependent Wightman functions \cite{Ema:2024hkj}.

Having established the reconstruction at the level of the linear kernel, we now apply it to the prepared gravitational response. The boundary traces below determine the combined regular pole profile, which is formed before physical strain conversion, as in Eq.~\eqref{eqS:converted-residue}. This order implements the construction for the noncompact gravitational problem. With matched higher-order data, causal support further extends it to the full waveform difference in Sec.~\ref{secS:time-inverse}.

\subsection{Coordinates and radiation traces}

We now reconstruct the intermediate field using its two radiation channels. In the Schwarzschild calculation we set $M=1$ and use
\begin{align}
 x=r_*&=r+2\log(r/2-1)~,~ \sigma=2/r~,\nonumber\\
 \tau&=t-H(r)~,\nonumber\\
 H(r)&=r+4\log(r/2)-2\log[(r-2)/2]~.
 \label{eqS:coordinates}
\end{align}
Infinity is at $\sigma=0$ and the horizon at $\sigma=1$. Since $H-x\to0$ at infinity and $H+x\to4$ at the horizon, the boundary times are $u=\tau$ and $v=\tau+4$, respectively. The canonical Laplace radiation amplitudes of a regular prepared field are therefore
\begin{equation}
 A_{\infty\beta}(s)=p_\beta(s,0)~,~
 A_{H\beta}(s)=e^{-4s}p_\beta(s,1)~.
 \label{eqS:boundary-traces}
\end{equation}
The horizon phase records the time offset between the two boundary conventions.

\subsection{Reconstruction of an isolated residue}
\label{secS:pole-reconstruction}

Take the residue of Eq.~\eqref{eqS:prepared-source} at $s_S$. Its source is analytic, so the residue solves the homogeneous radial equation. Let
\begin{equation}
 R_{p\beta}=\Res_{s=s_S}p_\beta~,~
 R_\beta(x)=e^{\ii\Om_SH(x)}R_{p\beta}(x)~.
\end{equation}
The second expression converts the regular hyperboloidal profile into its canonical radial profile.

Introduce advanced Jost solutions $u_A\sim e^{\ii\omega x}$ at the horizon and $v_A\sim e^{-\ii\omega x}$ at infinity. Continue these normalized solutions analytically to $\omega=\Om_S$, and define
\begin{equation}
 W_A=W(u_A,v_A)~,~
 K_H^A=-\frac{v_A}{W_A}~,~ K_\infty^A=-\frac{u_A}{W_A}~.
 \label{eqS:advanced-kernels}
\end{equation}
Primes in the Wronskian denote $\partial_x$. If $W_A(\Om_S)\ne0$, the two solutions form a basis. With $R_{b\beta}=\Res_{s=s_S}A_{b\beta}$, their endpoint Wronskians with the outgoing residue are
\begin{align}
 W(R_\beta,u_A)&=2\ii\Om_SR_{H\beta}~,\nonumber\\
 W(R_\beta,v_A)&=-2\ii\Om_SR_{\infty\beta}~.
\end{align}
Expanding $R_\beta$ in this basis then gives
\begin{equation}
 R_\beta(x)=2\ii\Om_S
 \sum_{b=H,\infty}K_b^A(x,\Om_S)R_{b\beta}~.
 \label{eqS:pole-reconstruction}
\end{equation}
Thus the two boundary residues determine the intermediate radial residue. The advanced profiles at complex frequency are defined by analytic continuation, rather than conjugation at that frequency.

Substituting Eq.~\eqref{eqS:pole-reconstruction} into the quadratic conversion gives
\begin{equation}
 K_{\alpha\beta}^{(S)}=\mathcal D_\alpha(z_S)
 \left[2\ii\Om_Se^{-\ii\Om_SH}
 \sum_{b=H,\infty}K_b^A R_{b\beta}\right]~.
 \label{eqS:converted-residue}
\end{equation}
The sum reconstructs the regular retarded profile on which $\mathcal D_\alpha$ acts. It is formed before conversion: the separate advanced profiles can have singular endpoint behavior outside the domain of the nonlinear map. This construction supplies the excitation in Eq.~\eqref{eqS:factorization} from lower-order radiation.

\subsection{Cancellation of the common cubic source}
\label{secS:waveform-difference}

The same idea extends to the full time dependence when two preparations share their first-order field and higher-order data prescription. Let $X_+$ and $X_-$ differ only in their second-order radiative initial data, and write
\begin{equation}
 \Delta\phi^{(2)}=\phi_{X_+}^{(2)}-\phi_{X_-}^{(2)}~.
\end{equation}
Subtracting Eq.~\eqref{eqS:perturbative-equations} yields
\begin{equation}
 \mathcal L\Delta\phi^{(2)}=0~,~
 \mathcal L\Delta\phi^{(3)}=B(\phi^{(1)},\Delta\phi^{(2)})~.
 \label{eqS:difference-equations}
\end{equation}
The common source $C(\phi^{(1)},\phi^{(1)},\phi^{(1)})$ cancels. Likewise, the purely first-order cubic term in Eq.~\eqref{eqS:observable-expansion} cancels, leaving
\begin{align}
 \Delta h_\alpha^{(3)}
 ={}&P_\alpha G_\alpha^R B_\alpha(\phi^{(1)},\Delta\phi^{(2)})
 \nonumber\\
 &+Q_\alpha(\phi^{(1)},\Delta\phi^{(2)})
 +P_\alpha\Delta\phi_{\rm data}^{(3)}~.
 \label{eqS:full-difference}
\end{align}
Here $\Delta\phi_{\rm data}^{(3)}$ is the homogeneous response associated with any change in third-order free data. The comparison in the main text fixes those data, so this last term vanishes.

In the numerical convention, the difference of the output master field and its velocity vanishes on the initial slice, and both strain primitives have fixed zero initial constants. These conditions fix the free radiative data while allowing the constrained metric components and nonlinear strain to follow from the field equations. Under a change of master variable, the initial-data prescription and the observable are transformed together.

Let $q$ be the master field associated with the homogeneous metric difference $\Delta\phi^{(2)}$. The source-dependent quadratic terms in its metric reconstruction cancel between preparations. Consequently the ordinary linear reconstruction takes $q$ to $\Delta\phi^{(2)}$. Once $q$ is determined by its radiation, every term on the right-hand side of Eq.~\eqref{eqS:full-difference} is fixed by quadratic dynamics.

\subsection{The time-dependent radiation inverse}
\label{secS:time-inverse}

We derive the radiation inverse explicitly. Extend the homogeneous master difference causally as $q_+=\Theta(\tau)q$. If $q_0=q(0)$ and $v_0=\dot q(0)$, distributional differentiation gives
\begin{equation}
 (\partial_\tau^2-L_2\partial_\tau-L_{14})q_+
 =q_0\delta'(\tau)+(v_0-L_2q_0)\delta(\tau)~.
 \label{eqS:slice-source}
\end{equation}
The initial preparation is thus represented by a source on the initial slice, with $L_{14}\equiv L_{1\ell}|_{\ell=4}$ the $\ell=4$ restriction of the spatial operator defined in Eq.~\eqref{eqS:hyperboloidal-operators}. In canonical coordinates, the Fourier-transformed source is
\begin{equation}
 \widetilde j_\Sigma=w e^{\ii\omega H}
 [(-\ii\omega-L_2)q_0+v_0]~,~
 w=1-H_x^2~.
 \label{eqS:canonical-slice-source}
\end{equation}
For Eq.~\eqref{eqS:coordinates}, $H_x=1-2\sigma^2$ and $w=4\sigma^2(1-\sigma^2)$.

Writing $\Psi_+(t,x)=q_+(t-H(x),x)$, the retarded Green operator propagates $j_\Sigma$ to the future. The advanced solution has support to the past of the same spacelike slice, so for $\tau>0$ the canonical field satisfies
\begin{equation}
 \Psi_+=(G^R-G^A)j_\Sigma
 \quad\hbox{after transformation to spacetime}~.
 \label{eqS:causal-support}
\end{equation}
This support property is the step that extends the pole argument to a complete waveform. At fixed frequency $G^A\widetilde j_\Sigma$ generally contributes; it disappears in the future only after the inverse transform.

For the canonical operator $-\partial_x^2+V_4^+-\omega^2$, we use the real-frequency scattering identity \eqref{eqS:scattering-identity} in the normalization of Eq.~\eqref{eqS:advanced-kernels}. Contracting the retarded boundary kernels with the slice source gives the future radiation amplitudes
\begin{align}
 \widetilde R_\infty(\omega)&=\int_0^\infty\dd\tau\,
 e^{\ii\omega\tau}\Delta\Psi_\infty^{(2)}(\tau)~,\nonumber\\
 \widetilde R_H(\omega)&=e^{4\ii\omega}\int_0^\infty\dd\tau\,
 e^{\ii\omega\tau}\Delta\Psi_H^{(2)}(\tau)~.
 \label{eqS:radiation-transforms}
\end{align}
Combining Eqs.~\eqref{eqS:causal-support} and \eqref{eqS:scattering-identity} then gives
\begin{align}
 q(\tau,x)=\int_{\mathbb R}\frac{\dd\omega}{2\pi}\,
 &e^{-\ii\omega[\tau+H(x)]}\,2\ii\omega\nonumber\\
 &\times\sum_{b=H,\infty}K_b^A(x,\omega)\widetilde R_b(\omega)~,
 \tau>0~.
 \label{eqS:time-reconstruction}
\end{align}
We denote this combined inverse by $\mathcal I_+$. Equations~\eqref{eqS:time-reconstruction} and \eqref{eqS:full-difference}, with fixed third-order data, are the reconstruction formulas of the main text.

For smooth compactly supported slice data, the source contractions are taken in their ordinary domain and the frequency integral in its Fourier limit. The linear scattering theory for the Schwarzschild radiative sector supplies the two-channel reconstruction~\cite{Xuan:2020wif,Masaood:2022bvi}. Extension to endpoint-reaching data requires convergence in a norm controlling the derivatives used in the quadratic conversion. The numerical examples below use smooth finite-energy data vanishing quadratically at both compactified endpoints and test this extension directly.

This inverse uses both future radiation histories and retains the transient response. Its isolated-pole limit recovers Eq.~\eqref{eqS:converted-residue}.

\section{Schwarzschild implementation and numerical results}
\label{secS:numerics}

\subsection{Evolution and strain normalization}

In the coordinates of Eq.~\eqref{eqS:coordinates}, the regular master equation is
\begin{equation}
 \ddot\Psi=L_{1\ell}\Psi+L_2\dot\Psi+b(\tau)~,
 \label{eqS:hyperboloidal}
\end{equation}
where
\begin{align}
 L_{1\ell}={}&\frac{\sigma^2(1-\sigma)}{16(1+\sigma)}\partial_\sigma^2
 +\frac{\sigma(2-3\sigma)}{16(1+\sigma)}\partial_\sigma
 -\frac{U_\ell}{16(1+\sigma)}~,\nonumber\\
 L_2={}&\frac{1-2\sigma^2}{4(1+\sigma)}\partial_\sigma
 -\frac{\sigma}{2(1+\sigma)}~,\nonumber\\
 U_\ell={}&\frac{8\lambda^2(\lambda+1)+12\lambda^2\sigma
 +18\lambda\sigma^2+9\sigma^3}{(2\lambda+3\sigma)^2}~.
 \label{eqS:hyperboloidal-operators}
\end{align}
Regular profiles at the characteristic endpoints implement the radiation boundary conditions. We discretize these operators at $N$ Chebyshev collocation points. The quadratic GR source and strain extraction follow Refs.~\cite{Bucciotti:2024zyp,Bucciotti:2024jrv}, using the hyperboloidal initial-value formulation of Ref.~\cite{Bourg:2025lpd}.

For a homogeneous even-parity mode, the linear strain conversion is
\begin{equation}
 h_{\ell m}=\alpha_\ell\Psi_{\ell m}~,~
 \alpha_\ell=\tfrac12\sqrt{(\ell+2)!/(\ell-2)!}~.
 \label{eqS:alpha}
\end{equation}
Let $a$ and $d$ denote the first-order strain amplitudes of the $220$ and $330$ modes. The intermediate $(4,4)$ master field is normalized by $(a/\alpha_2)^2$. Its interaction with the $220$ spectator produces the $(6,6)$ strain proportional to $a^3$, while the $330$ spectator produces the $(7,7)$ strain proportional to $a^2d$. The coefficients and waveforms below have these parent strain factors divided out. Complex positive-$m$ modes are used, with the reality-related negative-$m$ partners understood.

For a homogeneous intermediate field $v$, the conversion is implemented as
\begin{equation}
 \ddot Y_\alpha=L_{1\alpha}Y_\alpha+L_2\dot Y_\alpha
 +e^{s_D\tau}\mathcal B_{\alpha D}[v,\dot v]~.
 \label{eqS:forward-conversion}
\end{equation}
The exported source includes its time derivatives. Physical strain combines the outgoing $Y_\alpha$ with the quadratic reconstruction terms and both time primitives. These are the numerical realizations of the two terms in Eq.~\eqref{eqS:full-difference}. In the waveform comparison, only the even $(4,4)$ preparation varies. The other second-order channels remain common, so their cubic contributions, including the crossed $(5,5)$ route in the $(7,7)$ output, cancel.

\subsection{The isolated-pole calculation}
\label{secS:pole-numerics}

For the pole test, the production source is the regularized $220\times220\to(4,4)$ source $b e^{s_p\tau}$, with $s_p=2s_{220}$. Preparation $\beta=0$ sets the second-order master field and velocity to zero initially. Preparation $\beta=1$ adds the normalized seed
\begin{equation}
 p_{01}(\sigma)=\tfrac14\sigma^2(1-\sigma)^2~,~ v_{01}=0~.
 \label{eqS:pole-seed}
\end{equation}
These two preparations test the $440$ residue and are distinct from the pair $X_\pm$ used in Figs.~\ref{fig:prep} and \ref{fig:reconstruction}.

Writing $P=\mathcal L_4(s_p)^{-1}b$ separates the driven response:
\begin{align}
 p_\beta(\tau)&=P e^{s_p\tau}+v_\beta(\tau)~,\nonumber\\
 \ddot v_\beta&=L_{14}v_\beta+L_2\dot v_\beta~,\nonumber\\
 (v_\beta,\dot v_\beta)|_0&=(p_{0\beta}-P,v_{0\beta}-s_pP)~.
 \label{eqS:homogeneous-remainder}
\end{align}
We evolve this complete homogeneous remainder through Eq.~\eqref{eqS:forward-conversion} and extract the strain residue around $s_{440}+s_D$. The forward calculation obtains the intermediate excitation from the initial-value problem. The factorized prediction uses the separate quadratic conversion and boundary reconstruction.

The chosen contour isolates the required cubic coefficient. The driven source and genuine cubic forcing have frequency $s_p+s_D$, separated from the selected pole by $0.104/M$. The nearest computed output poles lie more than $0.09/M$ away. For the mixed $a^2d$ response, the nearest computed pole of the crossed $220+330\to(5,5)$ route is $0.0229/M$ away. Contour radii $0.002/M$, $0.004/M$, and $0.008/M$ exclude these singularities and the strain-primitive frequencies. The evolved sector therefore has the selected residue of the complete cubic response.

At $N=48$, the physical strain coefficients, with superscript $(440)$ suppressed, are
\begin{align}
 K_{66,0}&\simeq-0.098604-0.015125\ii~,\nonumber\\
 K_{66,1}&\simeq-0.097546-0.022317\ii~,\nonumber\\
 K_{77,0}&\simeq-0.095699-0.0063303\ii~,\nonumber\\
 K_{77,1}&\simeq-0.095281-0.013306\ii~.
 \label{eqS:gr-coefficients}
\end{align}
The separately evaluated quadratic conversions predict
\begin{equation}
 \frac{K_{77,\beta}}{K_{66,\beta}}
 =\frac{\mathcal V_{77|440,330}}{\mathcal V_{66|440,220}}
 \simeq0.95785-0.082729\ii~.
 \label{eqS:conversion-ratio}
\end{equation}
Changing the preparation changes each coefficient by about $7.3\%$. The unrounded residue matrix has singular-value ratio $9\times10^{-15}$; Table~\ref{tabS:pole-checks} reports its agreement with the factorized prediction. The two rows have different output frequencies and share the same intermediate pole after the spectator shifts are removed.

\begin{table}[tb]
\caption{Relative differences in the isolated-pole checks.}
\label{tabS:pole-checks}
\begin{ruledtabular}
\begin{tabular}{p{0.65\columnwidth}r}
Comparison & Difference\\
\hline
Forward residue versus factorization, $N=32,48,64$ & $2.6\times10^{-10}$\\
Boundary reconstruction, $N=32,48$ & $7.4\times10^{-9}$\\
Boundary reconstruction, including $N=64$ & $1.2\times10^{-7}$\\
Contour radius and sampling variation & $1.3\times10^{-11}$\\
Conversion-ratio test at $N=48$ & $2.0\times10^{-11}$\\
Time-fit window and basis variation & $3.5\times10^{-4}$
\end{tabular}
\end{ruledtabular}
\end{table}

The contours use 32--64 samples. The radiation calculation independently integrates the advanced Zerilli solutions, varying the integration step, matching position $x/M=50,60,80$, and asymptotic-series order $14,20,26$. Its largest discrepancy is $1.2\times10^{-7}$; conversion derivatives amplify radial errors at larger $N$.

Independent fourth-order evolution with steps $0.02M$, $0.01M$, and $0.005M$ gives successive waveform differences decreasing by a factor of approximately 16. Fits on $20\leq\tau/M\leq45$ recover the same residues. Varying the start from $20M$ to $25M$ and retaining two to four intermediate overtones, together with tail terms, gives the spread in Table~\ref{tabS:pole-checks}. The fit condition numbers are of order $10^9$--$10^{11}$.

\subsection{The preparations in Figs.~\ref{fig:prep} and \ref{fig:reconstruction}}
\label{secS:pair}

We next construct the two preparations whose quadratic radiation is nearly identical at infinity but whose cubic responses differ. Let $X=(q_0,v_0)$ denote homogeneous $(4,4)$ master data added to a common sourced second-order solution. The positive Zerilli Killing energy on the initial hyperboloidal slice is
\begin{align}
 E[X]=\int_0^1\dd\sigma\,\bigg[
 &\frac{\sigma^2(1-\sigma)}4|\partial_\sigma q_0|^2
 +\frac{U_4}4|q_0|^2\nonumber\\
 &+4(1+\sigma)|v_0|^2\bigg]~.
 \label{eqS:energy}
\end{align}
It follows from $\tfrac12\int\dd x\,[(1-H_x^2)|v_0|^2+|\partial_xq_0|^2+V_4^+|q_0|^2]$. This norm fixes the size of the added preparation.

We expand the initial field and velocity independently in
\begin{equation}
 \varphi_k(\sigma)=\sigma^2(1-\sigma)^2T_k(2\sigma-1)~,
 k=0,\ldots,p-1~,
 \label{eqS:trial-basis}
\end{equation}
where $T_k$ is a Chebyshev polynomial. If $c$ collects their coefficients, continuum quadrature gives $E=c^\dagger Gc$. Cholesky factorization $G=LL^\dagger$ and $c=L^{-\dagger}z$ then give $E=\|z\|^2$. This normalization is independent of the propagation grid.

Let $X_0$ be the unit-energy version of $(\varphi_0/4,0)$. A unit-energy direction $X_\delta$ is selected in the $p=18$ trial space from weakly radiating right-singular subspaces of the infinity observation map on $0\leq\tau/M\leq400$. Within each candidate subspace, the leading output singular vector selects a direction with appreciable cubic response; the candidates are compared under the radiation tolerance at the finer resolution. The resulting pair is
\begin{equation}
 X_\pm=\tfrac12X_0\pm a_\delta X_\delta~,~
 a_\delta\simeq0.3000~.
 \label{eqS:pair}
\end{equation}
Their energies are $0.32845$ and $0.35155$. The numerical archive specifies $a_\delta$ and $X_\delta$ at full precision in the basis of Eq.~\eqref{eqS:trial-basis}. The common first-order spectators and output-data prescription are fixed. The plotted boundary radiation is that of the added homogeneous fields, measured relative to the common sourced reference.

For a complex waveform, define
\begin{equation}
 \|f\|_{[a,b]}^2=\int_a^b\dd\tau\,|f(\tau)|^2~,
 \label{eqS:L2}
\end{equation}
with the sum over channels understood for a vector waveform. If $\bar R_\infty=(R_{\infty,+}+R_{\infty,-})/2$, the finely sampled records give
\begin{equation}
 \frac{\|R_{\infty,\pm}-\bar R_\infty\|_{[0,400M]}}
 {\|\bar R_\infty\|_{[0,400M]}}
 \simeq9.89\times10^{-7}~.
 \label{eqS:infinity-mismatch}
\end{equation}
The full pair separation is twice this mismatch, which is unchanged at the displayed precision for an $800M$ record. For $X_\delta$, almost all the energy exits through the horizon; the infinity flux is $2.8\times10^{-10}$.

The plotted cubic responses $h_{\alpha,X_\pm}^{(3)}$ share the same zero-seed subtraction. Their difference
\begin{equation}
 \delta h_\alpha^{(3)}
 =h_{\alpha,X_+}^{(3)}-h_{\alpha,X_-}^{(3)}
 \label{eqS:plotted-difference}
\end{equation}
equals the complete cubic difference between the preparations by Eq.~\eqref{eqS:full-difference}. The peak normalization used in the main text is
\begin{equation}
 H_\alpha=\max_{\tau,\,s=\pm}|h_{\alpha,X_s}^{(3)}(\tau)|~.
 \label{eqS:peak-scale}
\end{equation}
Figure~\ref{fig:prep} plots $100|\delta h_\alpha^{(3)}|/H_\alpha$; the peak values are listed in Table~\ref{tabS:pair-results}.

The corresponding separation in complex $L^2$ norm is
\begin{equation}
 d_\alpha=
 \frac{2\|\delta h_\alpha^{(3)}\|_{[0,70M]}}
 {\|h_{\alpha,X_+}^{(3)}\|_{[0,70M]}+
  \|h_{\alpha,X_-}^{(3)}\|_{[0,70M]}}~.
 \label{eqS:pair-separation}
\end{equation}
Table~\ref{tabS:pair-results} distinguishes this norm separation from the pointwise peak percentage.

The pair also quantifies a finite-precision reconstruction limit. An infinity record at the midpoint is compatible with either preparation at the tolerance in Eq.~\eqref{eqS:infinity-mismatch}. For any estimate $\widehat h$ based on that record, the triangle inequality gives
\begin{equation}
 \max_{s=\pm}
 \frac{\|\widehat h-h_s\|}{\|h_s\|}
 \geq\frac{\|h_+-h_-\|}{\|h_+\|+\|h_-\|}~.
 \label{eqS:ambiguity-bound}
\end{equation}
The numerical pair gives a $5.51\%$ lower bound for the combined baseline-subtracted cubic response on $[0,70M]$. It establishes finite-precision ambiguity for this preparation class and radiation record.

\subsection{Reconstruction of the cubic waveform difference}
\label{secS:waveform-numerics}

The calculation in Fig.~\ref{fig:reconstruction} reconstructs the same pair using both radiation channels. With the discretized state $\boldsymbol X=(q,\dot q)$, write
\begin{equation}
 \dot{\boldsymbol X}=\boldsymbol A_4\boldsymbol X~,~
 \boldsymbol R_j=\boldsymbol C e^{\boldsymbol A_4\tau_j}
 \boldsymbol X(0)~,
\end{equation}
where $\boldsymbol C$ extracts horizon and infinity master traces. A fixed trial map $\boldsymbol S_p$ represents the independent initial field and velocity using Eq.~\eqref{eqS:trial-basis}. Weighted radiation samples define an observation matrix $\boldsymbol O_N$, while response rows propagated from Eq.~\eqref{eqS:forward-conversion} define the physical output matrix $\boldsymbol F_N$.

We approximate the inverse of $\boldsymbol O_{48}$ by a filtered singular-value decomposition,
\begin{equation}
 \boldsymbol O_{48}=U\Sigma V^\dagger~,~
 J_f=V\,{\rm diag}[f(\sigma_j)]\,U^\dagger~.
 \label{eqS:filtered-inverse}
\end{equation}
The reconstructed coefficients are $\widehat z=J_f\boldsymbol R$, and the waveform prediction is $\widehat h=\boldsymbol F_{48}\widehat z$. The coefficient fit takes only the lower-order radiation as data.

The filter is selected before applying it to the pair, by comparing observation and output operators at $N=48$ and $64$. Specifically, truncated-SVD and Tikhonov filters are compared over 29 relative cutoffs between $10^{-9}$ and $10^{-2}$, minimizing
\begin{equation}
 \|\boldsymbol F_{48}J_f\boldsymbol O_{64}-\boldsymbol F_{64}\|
 +10^{-6}\|\boldsymbol O_{64}\|\,\|\boldsymbol F_{48}J_f\|~.
 \label{eqS:filter-choice}
\end{equation}
The operator norms use the energy-normalized trial space. The two terms balance resolution-dependent bias and radiation-noise amplification, independently of the generating coefficients and target waveform.

The plotted reconstruction uses $p=26$, giving 52 unknown coefficients, with both boundary records on $0\leq\tau/M\leq40$ sampled every $0.1M$. The selected Tikhonov filter is
\begin{equation}
 f(\sigma_j)=\frac{\sigma_j}{\sigma_j^2+(\eta\sigma_1)^2}~,
 \eta=10^{-4.5}~.
\end{equation}
The forward reference uses $N=64$ and direct state-vector evolution with step $0.00125M$. The predictor uses $N=48$; its output rows are evaluated with the same step on the finer $0.025M$ output grid. Both use the established quadratic GR sources and full strain extraction. The input records originate from the separately propagated reference pair.

By linearity, the difference of the two predictions is the prediction from $\Delta\Psi_H^{(2)}$ and $\Delta\Psi_\infty^{(2)}$. Define
\begin{equation}
 e_\alpha=
 \frac{\|\widehat{\delta h}_\alpha^{(3)}-\delta h_\alpha^{(3)}\|_{[0,70M]}}
 {\|\delta h_\alpha^{(3)}\|_{[0,70M]}}~.
 \label{eqS:waveform-error}
\end{equation}
Table~\ref{tabS:pair-results} gives the errors quoted in Fig.~\ref{fig:reconstruction}. The lower panels show $|\widehat{\delta h}_\alpha^{(3)}-\delta h_\alpha^{(3)}|/\max_\tau|\delta h_\alpha^{(3)}|$.

\begin{table}[tb]
\caption{Results for the preparations in the main text, using complex waveforms on $0\leq\tau/M\leq70$.}
\label{tabS:pair-results}
\begin{ruledtabular}
\begin{tabular}{p{0.48\columnwidth}cc}
Quantity & $(6,6)$ & $(7,7)$\\
\hline
Common peak $H_\alpha$ & $0.06815$ & $0.05681$\\
Peak $100|\delta h_\alpha^{(3)}|/H_\alpha$ & $8.11$ & $11.12$\\
$L^2$ separation $100d_\alpha$ & $9.58$ & $12.60$\\
Relative complex $L^2$ error $e_\alpha$ & $1.02\times10^{-5}$ & $7.30\times10^{-6}$\\
Maximum error / peak difference & $9.36\times10^{-5}$ & $6.45\times10^{-5}$
\end{tabular}
\end{ruledtabular}
\end{table}

\subsection{Resolution checks and the explicit scattering inverse}

The preparation-pair tests compare $N=32,48,64$, direct state-vector and response-row evolution, a halved integration step, and waveform samples spaced by $0.025M$ and $0.0125M$. The corresponding relative changes in the strain outputs or pair separation are below $10^{-9}$. Independent 80-, 160-, and 320-point quadratures verify the hidden direction's unit initial energy to better than $10^{-12}$.

The boundary-based inverse was also tested on a smooth complex wave packet with nonzero velocity, beyond the polynomial generating class of $X_\pm$. In the same parent-master units its data are
\begin{align}
 q_0&=\tfrac14\varphi_0(\sigma)
 \exp[-((\sigma-0.55)/0.22)^2+5\ii\sigma]~,\nonumber\\
 v_0&=0.06\varphi_0(\sigma)\cos(3\pi\sigma)e^{-2\ii\sigma}~.
\end{align}
Using $N=32$ and 26 trial functions per initial field with a $60M$ radiation record gives cubic errors $3.8\times10^{-8}$ and $3.9\times10^{-8}$. Increasing the inverse to $N=48$ and 34 functions reduces these to $1.4\times10^{-10}$ and $1.5\times10^{-10}$. The nonlinear strain terms contribute materially: omitting them gives approximately $8.0\%$ and $4.4\%$ errors in this test.

Equation~\eqref{eqS:time-reconstruction} is checked separately with independently integrated continuum Jost solutions and real-frequency synthesis. The test reconstructs the intermediate field at $x/M=-6,-2,2,6,10$ over $0.5\leq\tau/M\leq70$. A $409.6M$ radiation record and frequency cutoff $16/M$ give combined relative field error $9.9\times10^{-7}$ and maximum individual-radius error $1.2\times10^{-6}$. Increasing the exterior matching radius from $1200M$ to $1600M$ leaves the displayed accuracy unchanged. This tests the explicit scattering formula for the intermediate field; the cubic waveform results use the finite-dimensional observation inverse described above.

These noiseless tests use fixed quadratic GR operators and strain conventions. Finite radiation records and regularization set the inversion accuracy.

\section{A two-pole exchange model}
\label{secS:two-pole}

The rank constraint in Sec.~\ref{secS:rank} can be illustrated by combining two intermediate poles. We use the Schwarzschild family $44n$, production channels $(220,220)$ and $(320,320)$, and conversion channels $44n+220\to(6,6)$ and $44n+330\to(7,7)$. This model uses forward quadratic-response tables~\cite{Bucciotti:2024zyp,Bucciotti:2024jrv,BucciottiQuadraticQNMCode} and linear excitation factors~\cite{MotohashiQNEFData}.

For each channel, we remove the output response from the published strain ratio using a complex-contour Zerilli Wronskian. Denote the resulting production and conversion proxies by $\Pi_\beta$ and $\Gamma^{(2)}_{\alpha n}$. The assembled matrix is
\begin{equation}
 R^{(\mathcal I)}_{\alpha\beta}=
 \sum_{n\in\mathcal I}\Gamma^{(2)}_{\alpha n}
 \frac{Z_{44n}}{\Om_{B_\beta}+\Om_{C_\beta}-\Om_{44n}}\Pi_\beta~.
 \label{eqS:model-matrix}
\end{equation}
Each term is an outer product. Here $R^{(\mathcal I)}$ is a coefficient matrix assembled from several poles and parent pairs; $K^{(S)}$ in the main text is the physical residue at one selected pole across preparations. This distinction fixes the meaning of the two rank tests.

\begin{table}[b]
\caption{Rank diagnostics for the exchange model. The singular values are denoted by $\sigma_1\geq\sigma_2$.}
\label{tabS:rank-model}
\begin{ruledtabular}
\begin{tabular}{lcc}
Intermediate poles & $\sigma_2/\sigma_1$ & $\epsilon_{\rm rk}$\\
\hline
$440$ & $<10^{-15}$ & $<10^{-15}$\\
$440+441$ & $0.0161$ & $0.260$
\end{tabular}
\end{ruledtabular}
\end{table}

Table~\ref{tabS:rank-model} shows the departure from rank one when the first overtone is added, with $\|R^{(441)}\|_F/\|R^{(440)}\|_F\simeq1.20$. The rank statistic changes by less than $4\times10^{-11}$ when the complex-contour endpoint radius increases from $2000M$ to $5000M$. Unequal row and column normalizations make the singular-value ratio small, while the gain-invariant statistic retains the departure from rank one.

We also propagate this model causally from vanishing nonlinear initial data. Define
\begin{equation}
 F(\Om,\Om_A;t)=\frac{e^{-\ii\Om t}-e^{-\ii\Om_A t}}{\Om-\Om_A}~.
\end{equation}
It approaches $-\ii t e^{-\ii\Om_A t}$ as the frequencies coincide. With $\Om_T=\Om_B+\Om_C+\Om_D$, two sequential driven oscillators give
\begin{align}
 h^{\rm ex}_{\alpha\beta}(t)
 &=\sum_{n=0}^1R^{(n)}_{\alpha\beta}\mathcal T_{n,\alpha\beta}(t)~,\nonumber\\
 \mathcal T_{n,\alpha\beta}(t)
 &=F(\Om_T,\Om_A;t)-F(\Om_{44n}+\Om_D,\Om_A;t)~.
 \label{eqS:model-waveform}
\end{align}
Overall response conventions are absorbed into the coefficients. Four independent fits, including linear-output, prompt, and tail terms, recover the matrix to a relative error below $5\times10^{-15}$ within this prescribed waveform basis. Omitting $441$ leaves a $14\%$ maximum relative waveform residual at a zero-time fit start and shifts $\epsilon_{\rm rk}$ from $0.260$ to $0.070$. Extraction from a general cubic waveform also requires accounting for contact terms sharing the total-frequency basis.

\begin{figure*}[t]
\centering
\includegraphics[width=0.94\textwidth]{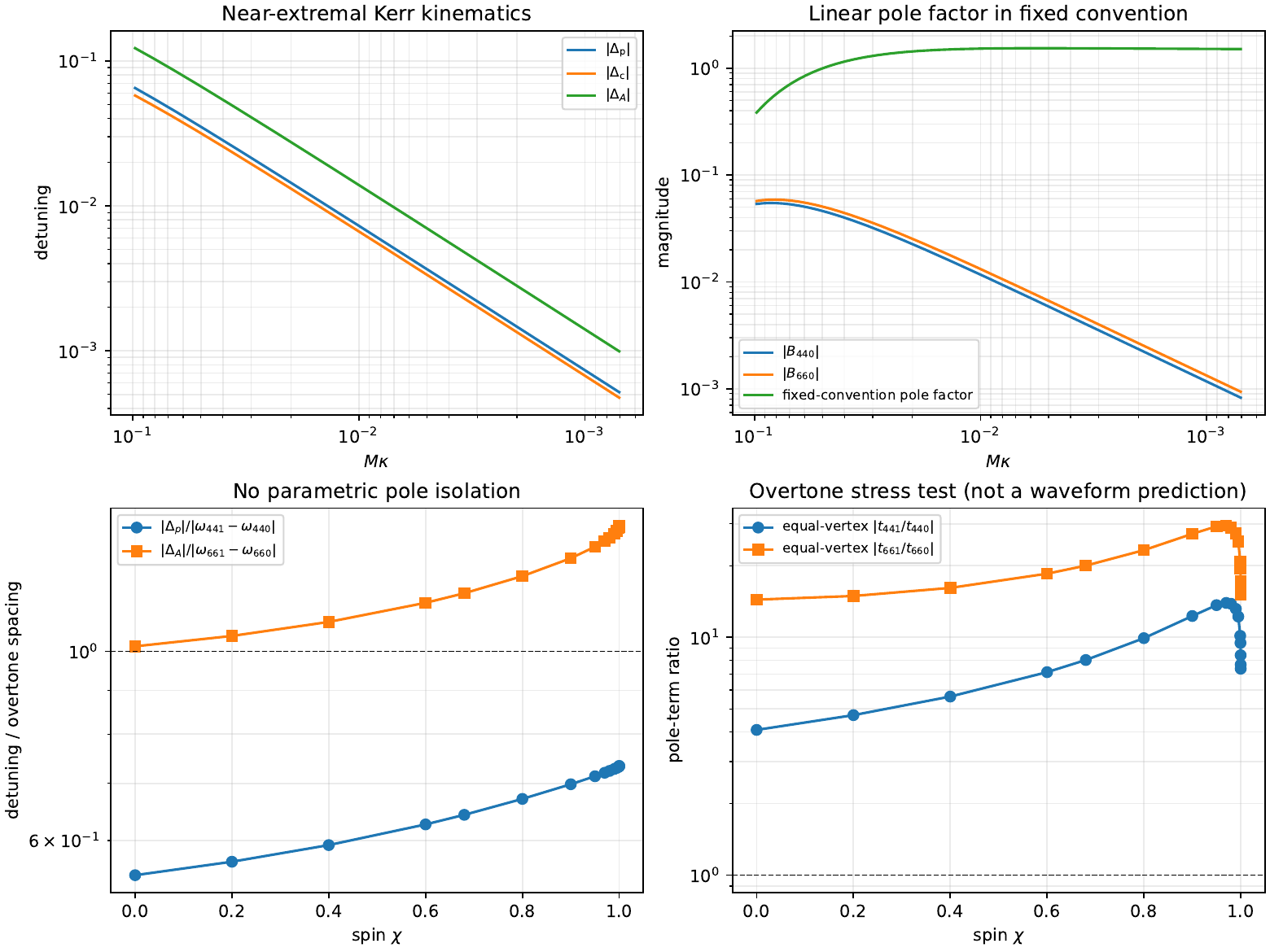}
\caption{Linear pole factors for the Kerr exchange chain. The detunings and excitation factors decrease together near extremality, leaving a finite combined linear factor. Ratios of detunings to overtone spacings remain of order unity. The equal-vertex overtone sums in the lower-right panel illustrate the effect of the pole tower; the physical sums require its nonlinear conversion and production coefficients.}
\label{figS:Kerr}
\end{figure*}

\section{Kerr pole factors near extremality}
\label{secS:Kerr}

Rotation introduces a closely spaced tower of corotating zero-damping modes as $\chi\to1$~\cite{Yang:2012pj,Berti:2006wq}. We examine the linear factors in the chain
\begin{equation}
 220+220\longrightarrow440~,~440+220\longrightarrow660~,
\end{equation}
using continuously tracked zero-damping branches. For $M=1$, define
\begin{align}
 \Delta_{\rm p}&=2\Om_{220}-\Om_{440}~,\nonumber\\
 \Delta_{\rm c}&=\Om_{440}+\Om_{220}-\Om_{660}~,\nonumber\\
 \Delta_A&=3\Om_{220}-\Om_{660}
 =\Delta_{\rm p}+\Delta_{\rm c}~.
 \label{eqS:Kerr-detunings}
\end{align}
Writing $\Om_I=m_I\Om_H+\kappa\lambda_I+O(\kappa^2)$, azimuthal conservation cancels the leading $m\Om_H$ terms at both vertices. All three detunings therefore scale as $\kappa$, where $\kappa$ is the surface gravity.

We use the public spin-$-2$ Teukolsky frequencies and excitation factors $B_n$~\cite{MotohashiQNEFData,Lo:2025njp}. Converting their $2M=1$ convention to $M=1$ divides the frequencies by two and the gravitational excitation factors by sixteen. An independent continued-fraction calculation with \texttt{qnm}~\cite{Stein:2019mop} agrees in frequency to better than $5\times10^{-11}$ over the tested spins. The excitation factors specify the linear pole normalization; the nonlinear vertices must be evaluated in that same convention.

\begin{table}[tb]
\caption{Exponents in $|X|\propto(M\kappa)^p$, fitted over $\chi\geq0.9999$ in the fixed Teukolsky convention.}
\label{tabS:Kerr}
\begin{ruledtabular}
\begin{tabular}{lc@{\qquad}lc}
$X$ & $p$ & $X$ & $p$\\
\hline
$\Delta_{\rm p}$ & $0.9974$ & $B_{440}$ & $1.0023$\\
$\Delta_{\rm c}$ & $0.9966$ & $B_{660}$ & $0.9999$\\
$\Delta_A$ & $0.9970$ & $K_0$ & $0.0077$
\end{tabular}
\end{ruledtabular}
\end{table}

The fundamental linear factor at the driven total frequency is
\begin{equation}
 K_0=\frac{B_{440}B_{660}}{\Delta_{\rm p}\Delta_A}~.
 \label{eqS:Kerr-factor}
\end{equation}
Table~\ref{tabS:Kerr} and Fig.~\ref{figS:Kerr} show that the excitation factors cancel the small denominators to leading power. At $\chi=0.999999$, $|K_0|\simeq1.51$. A full cubic amplitude also requires both nonlinear Kerr vertices in the same field normalization.

At the same spin, the isolation diagnostics are
\begin{align}
 \frac{|\Delta_{\rm p}|}{|\Om_{441}-\Om_{440}|}&\simeq0.734~,\nonumber\\
 \frac{|\Delta_A|}{|\Om_{661}-\Om_{660}|}&\simeq1.40~,\nonumber\\
 \frac{|\Delta_{\rm c}|}{|\Om_{661}-\Om_{660}|}&\simeq0.672~.
 \label{eqS:Kerr-isolation}
\end{align}
These ratios approach nonzero constants. The first two compare the independent propagator detunings with the corresponding overtone spacings; the third describes the conversion mismatch. Thus near-extremality brings neighboring intermediate and output poles into the same frequency scale as the detunings, motivating a collective treatment.

A quantitative Kerr rank prediction requires the nonlinear production and conversion vectors in addition to these pole factors.

\bibliography{ref}
\bibliographystyle{apsrev4-2}

\end{document}